\documentclass[12pt]{article}
\usepackage{graphics}
\usepackage{epsfig}
\usepackage{color}
\usepackage{amsmath}
\usepackage{amssymb}
\def\xxx#1 {{\sf hep-th/#1} }

\def\D{\Delta}

\def\a{\alpha}

\def\d{\delta}
\def\e{\varepsilon}

\def\m{\mu}
\def\tr{\mbox{tr}}

\def\s{\sigma}
\def\r{\rho}
\def\l{\lambda}
\def\t{\tau}
\def\o{\omega}

\def\vt{\vartheta}
\def\mc{\mathcal}

\def\p{\partial}

\def\ra{\rangle}
\def\dg{\dagger}

\numberwithin{equation}{section} 
\parskip=\medskipamount

\def\Nop{\bbbn}

\def\bbbn {{\rm I\!N}}
\def\RR{R-R }

\begin{document}

\thispagestyle{empty}
\def\thefootnote{\fnsymbol{footnote}}
\begin{flushright}
  hep-th/0304114 \\
  SPIN-2003/09 \\
  ITP-2003/16
\end{flushright}

\vskip 0.5cm

\begin{center}\LARGE
{\bf Open String Plane-Wave Light-Cone Superstring Field Theory}
\end{center}

\vskip 1.0cm

\begin{center}
{\large B. Stefa\'nski, jr.\footnote{E-mail address: {\tt stefansk@phys.uu.nl}}}

\vskip 0.5cm

{\it Spinoza Institute, University of Utrecht \\
Postbus 80.195, 3508 TD Utrecht, The Netherlands}
\end{center}

\vskip 1.0cm

\begin{center}
April 2003
\end{center}

\vskip 1.0cm

\begin{abstract}
\noindent 
We construct the cubic interaction vertex and dynamically
generated supercharges in light-cone superstring field theory for a large class
half-supersymmetric D-branes in the plane-wave background. We show that these satisfy
the plane-wave superalgebra at first order in string coupling. 
The cubic interaction vertex and dynamical supercharges presented here are given explicitly
in terms of oscilators and can be used to compute three-point functions of open
strings with endpoints on half-supersymmetric D-branes.
\end{abstract}

\vfill

\setcounter{footnote}{0}
\def\thefootnote{\arabic{footnote}}
\newpage

\renewcommand{\theequation}{\thesection.\arabic{equation}}
\section{Introduction}
\setcounter{equation}{0}

\noindent Because of their non-perturbative, yet quantifiable nature, D-branes~\cite{PolRR} 
have played a central role in many string theory settings. Following the discovery of a new exact
string theory background~\cite{bfhp,m,mt} - the maximally supersymmetric plane-wave of the Type IIB theory - it is natural to investigate D-branes in this spacetime.

\noindent In light-cone gauge the superstring action in the plane-wave background is quadratic and has been quantised~\cite{m,mt}. The 
spectrum of closed string states consists of a unique massless groundstate and an infinite tower of 
excited states whose masses are of order $\omega_n/(\alpha^\prime p^+)$ where
\begin{equation}
\omega_n\equiv\sqrt{n^2+(\alpha^\prime\mu p^+)^2}\,,\qquad n\in\Nop\,,
\end{equation}
$p^+$ is momentum along the $x^-$ direction and $\mu$ is the \RR flux. 
Because the action is quadratic, it has been possible to analyse D-branes in this background through the open strings that end on them~\cite{dp,st1,st2}. In this approach worldsheets with boundaries in $\sigma$ are considered; consistent
D-branes can then be thought of as boundary conditions on fields which are compatible with the variational principle. Since the closed string spectrum is known explicitly, it is also possible to investigate D-branes as sources for closed string states. In this approach boundary conditions
for closed string states are enforced at constant $\tau$, and are implemented, at the
Fock space level, by coherent states~\cite{bp,bgg,bpz,gg,st2}. The two approaches are compatible via the open-closed string duality of the cylinder diagram; this consistency condition is often refered to as the Cardy condition and has been explicitly verified for a large class of D-branes in the plane-wave background~\cite{bgg,gg} where a number of beautiful identities were proved. Given this understanding of D-branes, their interactions in this background merit further investigation.

\noindent In light-cone gauge in flat spacetime the action has a classical conformal symmetry which can be exploited to use a vertex operator approach to interactions. This symmetry is lost in the plane-wave background and the study of string interactions becomes much more difficult. The light-cone string field theory formalism has been succesfully used to study closed string interactions in the plane-wave background~\cite{sv,sv2,ps}. This formalism was originaly developed in flat spacetime for the bosonic string~\cite{Mand,CG,KK} in which a cubic interaction vertex for the scattering of three strings is constructed by requiring continuity of string fields on the worldsheet.
This continuity is enforced by a delta function on string fields
\begin{equation}\label{deltfn}
\Delta(X_1+X_2-X_3)\,.
\end{equation}
\begin{figure}[htb]
\begin{center}
\vspace{1cm}
\includegraphics{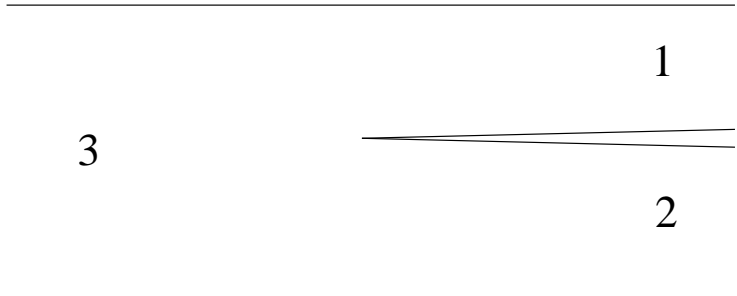}
\end{center}
\caption{The worldsheet of the cubic open string interaction vertex.}
\label{fig1}
\end{figure}
It is particularily useful to express the above functional delta function in Fourier modes and to construct a vertex as an exponential of Fock space creation operators which enforce the above
continuity conditions mode by mode~\cite{CG}. The formalism 
was subsequently extended to the superstring in flat spacetime~\cite{mandsusy,gs,gsb,gssft}. In this approach the
cubic interaction vertex is a first order correction in string coupling $g_s$ to the free Hamiltonian.
In order to satisfy the superalgebra with this new Hamiltonian, dynamical supercharges also receive corections. As a result, in the oscilator basis, the vertex (and the dynamical supercharges) is no
longer given by an exponential of Fock space creation operators as in the bosonic case. Instead it also contains a prefactor polynomial in creation operators. These prefactors are such that they do not destroy worldsheet continuity enforced by the exponential part of the vertex. They also ensure
that the superalgebra is satisfied order by order in the string coupling.

\noindent In this paper we will construct the cubic
interaction vertex and dynamical supercharges for open strings in the plane-wave background ending on D$_{-}$-branes. These constitute a large class of half-supersymmetric D-branes in this background. They extend along $q=0,\dots,4$ directions in $SO(4)$ and along $q\pm 2$ directions along $SO(4)^\prime$,
as well as, in the open string light-cone gauge, along the light-cone directions. In the bulk of the paper we will take the D-branes to lie at $x^{r^\prime}=0$ in the transverse space. It is, of course, possible to move such D$_-$-branes away from the origin using the isometries that they 
preserve~\cite{st1}. The cubic vertex and dynamical supercharges for such D$_-$-branes can be easily
obtained, by applying the aforementioned isometries to the vertex and supercharges computed explicitly below. 

\noindent The interaction vertex presented below uses the $SU(4)$ formalism developed in flat spacetime in~\cite{gssft}. In~\cite{gssft} the $SO(8)$ symmetry of the problem was broken to a $SO(6)\times SO(2)\sim SU(4)\times U(1)$ subgroup. In the $SO(8)$ formalism the spinor $S_a$ is its own conjugate momentum which introduces constraints that can significantly complicate the problem. The $SU(4)$ formalism circumvents this problem since an $SO(8)$ spinor breaks into two fields, one of which can be treated as the coordinate and the other as the conjugate momentum. We extend the $SU(4)$  formalism to open strings ending on D$_-$-branes in the plane-wave background. 
While it is true that the D$_-$-branes do not preserve the $SU(4)\times U(1)$ symmetry, the cubic vertex can still be written using $SU(4)$ notation. This is analogous to the closed string vertex in the plane-wave background which is written in $SO(8)$ notation even though the backgorund only preserves a $SO(4)\times SO(4)$ subgroup.

\noindent Following its discovery, the plane-wave background has received substantial interest within the AdS/CFT 
correspondence~\cite{malda,GKP,Witten}. It was conjectured~\cite{bmn} that string theory in this background is dual to a double scaling limit of $SU(N)$, ${\cal N}=4$ Super Yang-Mills theory. Subsequently, the duality has been extensively tested and investigated~\cite{bmnall,bits}.
Since D-branes are classical vacua of string field theory, one may expect that they will have to manifest themselves in the dual gauge theory. Several authors investigated gauge theories dual to D-branes in the plane-wave background~\cite{bgmnn,lp,st0}. A general picture was proposed in~\cite{st0}, where it was argued that the D-branes in the plane-wave background would be dual to suitable limits of the $SU(N)$, ${\cal N}=4$ Super Yang-Mills theory coupled to a defect CFT~\cite{kr1,kr2,dfo,egk}. The cubic vertex constructed in this paper can be used to investigate this correspondence at the level of open-string interactions.

\noindent While this paper was in the final stages of preparation~\cite{CK} appeared on the archive  in which functional expressions in $(SU(2)\times U(1))^2$ notation are given for the cubic interaction vertex for the D7-brane in the plane wave background. 

\noindent This paper is organized as follows. In section~\ref{sec2} we present
the free open string in the plane-wave background and set out the $SU(4)$ notation. 
In sections~\ref{sec3} we construct the open superstring three point vertex in the 
plane-wave background. An appendix is included which contains some of the computational 
details.
\section{The free open string theory on the plane-wave in the $SU(4)$ formalism}\label{sec2}

In the $SU(4)$ formalism the $SO(8)$ vector representation decomposes as
\begin{equation}
{\bf 8}_v={\bf 6}_0+{\bf 1}_1+{\bf 1}_{-1}\,,
\end{equation}
where the subscripts denote $U(1)$ quantum numbers. The ${\bf 8}_v$ index $I=1,\dots,8$ 
breaks into $i=1,\dots,6$ of the ${\bf 6}$, as well as $L$ and $R$ of the two singlets. 
We will take $i$ to lie along the $X^1,\dots,X^6$ directions and 
\begin{equation}
X^R=\frac{1}{\sqrt{2}}\left(X^7+iX^8\right)\,,\qquad
X^L=\frac{1}{\sqrt{2}}\left(X^7-iX^8\right)\,.
\label{LRdef}
\end{equation}
The spinor representations decompose as
\begin{equation}
{\bf 8}_s={\bf 4}_{1/2}+{\bf {\bar 4}}_{-1/2}\,,\qquad 
{\bf 8}_c={\bf 4}_{-1/2}+{\bf {\bar 4}}_{1/2}\,.
\end{equation}
The ${\bf 4}$ and ${\bf{\bar 4}}$ indices will be denoted by a superscript and subscript 
$A=1,\dots,4$ index, respectively.
An open string ending on a half-supersymmetric D-branes in the plane-wave background in
light-cone gauge is described, in the $SU(4)$ formalism by 
$x^i_u(\s_u)$, $x^L_u(\s_u)$, $x^R_u(\s_u)$,   and $\lambda^A_u(\s_u)$, $\theta_{uA}(\s_u)$,
${\tilde \lambda}^A_u(\s_u)$, ${\tilde \theta}_{uA}(\s_u)$ 
\footnote{We will often suppress these indices in what follows.} together with
a set of boundary conditions. These have been discussed at length in~\cite{st1}. For D-branes presently under consideration these are
\begin{eqnarray}
\partial_\sigma x^r_u|_{\sigma=0,\pi|\alpha_u|}&=&0\,,\qquad\qquad\qquad\qquad
x^{r^\prime,L,R}_u|_{\sigma=0,\pi|\alpha_u|}=0\\
\theta_{uA}|_{\sigma=0,\pi|\alpha_u|}
&=&\Omega_A{}^B{\tilde\theta}_{uB}|_{\sigma=0,\pi|\alpha_u|}\,,\qquad
\lambda^A_u|_{\sigma=0,\pi|\alpha_u|}=
\Omega^A{}_B{\tilde\lambda}^{B}_u|_{\sigma=0,\pi|\alpha_u|}\,.
\end{eqnarray}
Above the superscripts $r$ and $r^\prime$, which together span $i$, correspond to Neumann and Dirichlet
boundary conditions respectively; the matrix $\Omega$ is defined as~\footnote{We are using the same gamma-matrix conventions as~\cite{gssft}.}
\begin{equation}
\Omega=\rho^{RL}\prod_{r^\prime}\rho^{r^\prime}\,.
\end{equation}
Recall that for D$_-$-branes $\Omega\Pi\Omega\Pi=-1$ with $\Pi=\rho^{1234}$.
The index $u=1,2,3$ denotes the $u$th string (see Figure~\ref{fig1}). All the D-branes
under consideration here have at least two Dirichlet directions; we take $L,R$ to be
Dirichlet througout.

\noindent In a collision process $p^+_u$ will be positive for an incoming string and
negative for an outgoing one and it is convenient to define
\begin{equation}
\alpha_u=\alpha^\prime p^+_u\,.
\end{equation}
The mode expansions of the bosonic fields at $\t=0$ are
\begin{eqnarray}
x^r_u(\sigma_u)&=&x^r_{0(u)}+\sqrt{2}\sum_{n=1}^\infty x^r_{n(u)}
\cos\frac{n\sigma_u}{|\alpha_u|}\,,\\
p^r_u(\sigma_u)&=&\frac{1}{\pi|\alpha_u|}[p^r_{0(u)}+\sqrt{2}\sum_{n=1}^\infty p^r_{n(u)}
\cos\frac{n\sigma_u}{|\alpha_u|}]\,,\\
x^{r^\prime}_u(\sigma_u)&=&\sqrt{2}\sum_{n=1}^\infty x^{r^\prime}_{-n(u)}
\sin\frac{n\sigma_u}{|\alpha_u|}\,,\\
p^{r^\prime}_u(\sigma_u)&=&\frac{\sqrt{2}}{\pi|\alpha_u|}\sum_{n=1}^\infty p^{r^\prime}_{-n(u)}\sin\frac{n\sigma_u}{|\alpha_u|}\,,
\end{eqnarray}
with the $L,R$ directions having the same expansions as the $r^\prime$. The Fourier modes can be re-expressed in terms of creation and annihilation operators as
\begin{equation}\label{xp}
x_{n(u)}^I=i\sqrt{\frac{\a'}{2\o_{n(u)}}}\bigl(a_{n(u)}^I-a_{n(u)}^{I\,\dg}\bigr)\,,\qquad
p_{n(u)}^I=\sqrt{\frac{\o_{n(u)}}{2\a'}}\bigl(a_{n(u)}^I+a_{n(u)}^{I\,\dg}\bigr)\,,\qquad
\end{equation}
with $n$ non-negative for $I=r$ and negative for $I=r^\prime,\,L,\,R$.
Canonical quantization of the bosonic coordinates
\begin{equation}
[x_r^I(\s_u),p_s^J(\s_v)]=i\d^{IJ}\d_{uv}\d(\s_u-\s_v)
\end{equation}
implies the commutation relations
\begin{equation}
[a_{n(u)}^I,a_{m(v)}^{J\,\dg}]=\d^{IJ}\d_{nm}\d_{uv}\,.
\end{equation}

\noindent The following linear combinations of the fermionic fields are the fermionic normal modes
\begin{eqnarray}\label{normlambda}
\lambda^A_{\pm u}(\sigma_u)&=&
\lambda^A_u(\sigma_u)\pm\Omega^A{}_B{\tilde\lambda}^B_u(\sigma_u)\,,\\ \label{normtheta}
\theta_{\pm uA}(\sigma_u)&=&
\theta_{uA}(\sigma_u)\pm\Omega_A{}^B{\tilde\theta}_{uB}(\sigma_u)\,.
\end{eqnarray}
At $\tau=0$ the mode expansions of $\lambda^A_{\pm(u)}$ and $\theta_{\pm(u)A}$ are
\begin{eqnarray}
\lambda^A_{+u}(\sigma_u)&=&\frac{1}{\pi|\alpha_u|}
\left[\sqrt{2}\lambda_{0(u)}^A+2\sum_{n=1}^\infty
\lambda_{n(u)}^A\cos\frac{n\sigma_u}{|\alpha_u|}\right]\,,\\
\lambda^A_{-u}(\sigma_u)&=&\frac{1}{\pi|\alpha_u|}
\sum_{n=1}^\infty \lambda^A_{-n(u)}
\sin\frac{n\sigma_u}{|\alpha_u|}\,,\\
\theta_{+uA}(\sigma_u)&=&\sqrt{2}\theta_{0(u)A}+2\sum_{n=1}^\infty
\theta_{n(u)A}\cos\frac{n\sigma_u}{|\alpha_u|}\,,\\
\theta_{-uA}(\sigma_u)&=&2\sum_{n=1}^\infty
\theta_{-n(u)A}\sin\frac{n\sigma_u}{|\alpha_u|}\,.
\end{eqnarray}
The non-zero anti-commutators are
\begin{equation}
\left\{\theta_{\pm A u}(\sigma_u),\lambda_{\pm v}^B(\sigma^\prime_v)\right\}=
2\delta_{uv}\delta_A^B\delta(\sigma_u-\sigma^\prime_v)\,,
\end{equation}
or in terms of modes
\begin{equation}
\left\{\theta_{m(u)A},\lambda_{n(v)}^B\right\}=\delta_{mn}\delta_{uv}\delta_A^B\,.
\end{equation}
It is convenient to define a new basis of non-zero moded oscilators 
\begin{eqnarray}
\lambda_{n(u)}^A&=&\sqrt{\frac{|n|}{2\omega_{n(u)}}}
\left(P^{-1}_{n(u)C}{}^AR^C_{n(u)}+P_{n(u)C}{}^AR^C_{-n(u)}\right)\,,\\
\lambda_{-n(u)}^A&=&i\sqrt{\frac{|n|}{2\omega_{n(u)}}}
\left(P_{n(u)C}{}^AR^C_{n(u)}-P^{-1}_{n(u)C}{}^AR^C_{-n(u)}\right)\,,\\
\theta_{n(u)A}&=&\sqrt{\frac{|n|}{2\omega_{n(u)}}}
\left(P_{n(u)A}{}^CR_{n(u)C}+P^{-1}_{n(u)A}{}^BR_{-n(u)B}\right)\,,\\
\theta_{-n(u)A}&=&i\sqrt{\frac{|n|}{2\omega_{n(u)}}}
\left(P^{-1}_{n(u)A}{}^CR_{n(u)C}-P_{n(u)A}{}^BR_{-n(u)B}\right)\,.
\end{eqnarray}
Here
\begin{equation}
P^{\pm 1}_{n(u)A}{}^B=\frac{1}{\sqrt{1-\rho_{n(u)}^2}}(1\mp i\rho_n\Omega\Pi)_A{}^B\,,
\end{equation}
is the generalisation of the $P^{\pm 1}_{n(r)}$ in the closed string and
\begin{equation}
\r_{n(r)}=\r_{-n(r)}=\frac{\o_{n(r)}-|n|}{\m\a_r}\,,\qquad
c_{n(r)}=c_{-n(r)}=\frac{1}{\sqrt{1+\r_{n(r)}^2}}\,.
\end{equation}
These modes satisfy 
\begin{equation}
\{R_{An(u)},R^B_{m(v)}\}=\d_A^B\d_{n+m}\d_{uv}\,.
\end{equation}

\noindent The free string light-cone Hamiltonian is
\begin{eqnarray}
H_{2(u)}&=&\int_0^{\pi|\alpha_u|}d\sigma_u\left[
\frac{e(\alpha_u)}{2\pi\alpha^\prime}(\pi^2\alpha^\prime{}^2(p^I_u)^2+
(\partial_\sigma x^I_u)^2+\mu^2(x^I_u)^2)\right.
\nonumber\\
&&\left.+\frac{1}{2 i}(e(\alpha_u)(\theta_{+uA}\lambda^{\prime A}_{-u}+
\theta_{-uA}\lambda^{\prime A}_{+u})+\mu(\lambda_{+u}^A(\Omega\Pi)_A{}^B\theta_{+uB}-
\lambda_{-u}^A(\Omega\Pi)_A{}^B\theta_{-uB}))\right]\,.
\end{eqnarray}
It is convenient to define
\begin{equation}
R_{\pm 0(u)}^A=\frac{1}{2}(1\mp i e(\alpha_u)\Omega\Pi)^A{}_B\lambda_{0(u)}^B\,,\qquad
R_{\pm 0(u)A}=\frac{1}{2}(1\mp i e(\alpha_u)\Omega\Pi)_A{}^B\theta_{0(u)B}\,,
\end{equation}
which satisfy
\begin{equation}
\left\{R_{\pm 0A(u)},R_{\pm 0(v)}^B\right\}=0\,,\qquad
\left\{R_{\pm 0A(u)},R_{\mp 0(v)}^B\right\}=\delta_{uv}\frac{1}{2}(1\mp i\Omega\Pi)_A{}^B\,,
\end{equation}
The mode-expanded Hamiltonian is then
\begin{eqnarray}
H_{2(u)}&=&\frac{1}{\a_u}\sum_{n=1}^\infty
\o_{n(u)}\bigl(a_{n(u)}^{r\dg}a^r_{n(u)}+a^{r^\prime\dg}_{n(u)}a^{r^\prime}_{n(u)}
+R_{-n(u)A}R_{n(u)}^A+R^A_{-n(u)}R_{An(u)}\bigr)\nonumber \\
&&+\frac{\o_{0(u)}}{\alpha_u}(a_{0(u)}^{r\dg}a^r_{0(u)}+R_{-0(u)A}R^A_{+0(u)}
+R_{-0(u)}^AR_{+0(u)A})+\frac{1}{2}\mu e(\alpha_u)(p-5)\,,
\end{eqnarray}
where the last term is the zero-mode normal ordering constant for a D$p_-$-brane. The vacuum $|v\ra_u$ is defined as~\footnote{This vacuum is the $SU(4)$ version of the vacuum defined in~\cite{st2}.}
\begin{equation}
a^I_{n(u)}|v\ra_u=0\,,\qquad R_{m(u)}^A|v\ra_u=0\,,\qquad R_{m(u)A}|v\ra_u=0\,,\quad m > 0
\end{equation}
with $n$ negative (non-negative) for $I$ Dirichlet (Neumann) and for 
\begin{equation}
R_{+0(u)}^A|v\ra_u=0\,,\qquad R_{+0(u)A}|v\ra_u=0\,.
\end{equation}
The 16 supersymmetries are generated by the kinematical $q^+_{(u)A}$, $q^{+A}_{(u)}$ and the
dynamical $q^-_{(u)A}$, $q^{-A}_{(u)}$. The superalgebra was computed explicitly in~\cite{st2}.
In the $SU(4)$ formalism the part of the superalgebra of importance throughout this paper is
\begin{equation}\label{sualg}
\left\{q^{-A}_{(u)},q^-_{(u)B}\right\}=2\delta^A{}_B\delta_{uv}H_{2(u)} + i\mu\delta_{uv}\left((\gamma^{ij}\Pi){}^A{}_B J^{ij}_{(u)}+(\gamma^{i^\prime j^\prime 
}\Pi^\prime){}^A{}_B  J^{i^\prime j^\prime}_{(u)} -{\tilde\Pi}^A{}_BJ^{LR}_{(u)}
\right)\,,
\end{equation}
where $\Pi^\prime=\rho^{5678}$ and ${\tilde\Pi}=\Pi,\Pi^\prime$ for $R,L$ in $SO(4)$ or 
$SO(4)^\prime$, respectively.
The non-linearly realized supercharges can be expressed in terms of the normal modes of the system as
\begin{eqnarray}
q^-_{(u)A}&=&\int_0^{\pi|\alpha_u|}d\sigma_u\Bigl[
-e(\alpha_u)(\pi\alpha^\prime p^r_u\rho^r_{AB}\lambda^B_{+u}-\partial_\sigma x^r_u\rho^r_{AB}\lambda^B_{-u})
-\mu x^r_u(\rho^r\Omega\Pi)_{AB}\lambda^B_{+u}\nonumber \\ &&
-e(\alpha_u)(\pi\alpha^\prime p^{r^\prime}_u\rho^{r^\prime}_{AB}\lambda^B_{-u}
-\partial_\sigma x^{r^\prime}_u\rho^{r^\prime}_{AB}\lambda^B_{+u})
+\mu x^{r^\prime}_u(\rho^{r^\prime}\Omega\Pi)_{AB}\lambda^B_{-u}\nonumber \\ &&
+\sqrt{2}(\alpha^\prime p^R_u\theta_{-uA}-\frac{1}{\pi}\partial_\sigma x^R_u\theta_{+uA}-
e(\alpha_u)\frac{\mu}{\pi}x^R_u(\Omega\Pi)_A{}^B\theta_{-uB})\Bigr]\,,\\
q^{-A}_{(u)}&=&\int_0^{\pi|\alpha_u|}d\sigma_u\Bigl[
(\alpha^\prime p^r_u\rho^{rAB}\theta_{+uB}-\frac{1}{\pi}\partial_\sigma x^r_u\rho^{rAB}\theta_{-uB}
+e(\alpha_u)\frac{\mu}{\pi} x^r_u(\rho^r\Omega\Pi)^{AB}\theta_{+uB})\nonumber \\ &&
+(\alpha^\prime p^{r^\prime}_u\rho^{r^\prime AB}\theta_{-uB}-
\frac{1}{\pi}\partial_\sigma x^{r^\prime}_u\rho^{r^\prime AB}\theta_{+uB}
-e(\alpha_u)\frac{\mu}{\pi} x^{r^\prime}_u(\rho^{r^\prime}\Omega\Pi)^{AB}\theta_{-uB})\nonumber \\ &&
+\sqrt{2}e(\alpha_u)(\pi\alpha^\prime p^L_u\lambda^A_{-u}-\partial_\sigma x^L_u\lambda^A_{+u})-\sqrt{2}\mu x^L_u(\Omega\Pi)^A{}_B\lambda^B_{-u})\Bigr]\,.
\end{eqnarray}
Expanding in modes one finds
\begin{eqnarray}
q^-_{(u)A}&=&-\sqrt{2}\frac{\alpha^\prime}{\alpha}p_{0(u)}^r(\rho^r\lambda_{0(u)})_A-\sqrt{2}\mu x^r_{0(u)}(\rho^r\Omega\Pi\lambda_{0(u)})_A
\nonumber \\&&
-\frac{2}{\alpha_u}\sqrt{\frac{\alpha^\prime}{2}}\sum_{n=1}^\infty\sqrt{n}\Bigl[
a^r_{n(u)}(\rho^rP_{n(u)}R_{-n(u)})_A+a^{r\dagger}_{n(u)}(\rho^rP^{-1}_{n(u)}R_{n(u)})_A
-ia^{r^\prime}_{n(u)}(\rho^{r^\prime}P^{-1}_{n(u)}R_{-n(u)})_A\nonumber\\&&
+ia^{r^\prime\dagger}_{n(u)}(\rho^{r^\prime}P_{n(u)}R_{n(u)})_A
+i\sqrt{2}\alpha_u a^R_{n(u)}(P_{n(u)}R_{-n(u)})_A-i\sqrt{2}\alpha_u a^{R\dagger}_{n(u)}(P^{-1}_{n(u)}R_{n(u)})_A\Bigr]\\
q^{-A}_{(u)}&=&\sqrt{2}\alpha^\prime p_{0(u)}^r(\rho^r\theta_{0(u)})^A+\sqrt{2}\mu\alpha_u x^r_{0(u)}(\rho^r\Omega\Pi\theta_{0(u)})^A
\nonumber \\&&
+\sqrt{2\alpha^\prime}\sum_{n=1}^\infty\sqrt{n}\Bigl[
a^r_{n(u)}(\rho^rP^{-1}_{n(u)}R_{-n(u)})^A+a^{r\dagger}_{n(u)}(\rho^rP_{n(u)}R_{n(u)})^A
-ia^{r^\prime}_{n(u)}(\rho^{r^\prime}P_{n(u)}R_{-n(u)})^A\nonumber\\&&
+ia^{r^\prime\dagger}_{n(u)}(\rho^{r^\prime}P^{-1}_{n(u)}R_{n(u)})^A
-i\frac{\sqrt{2}}{\alpha_u} a^L_{n(u)}(P^{-1}_{n(u)}R_{-n(u)})^A+i\frac{\sqrt{2}}{\alpha_u} a^{L\dagger}_{n(u)}(P_{n(u)}R_{n(u)})^A\Bigr]
\end{eqnarray}
\section{The cubic open string interaction}\label{sec3}

In this section we construct the cubic open string interaction vertex, corresponding to two strings joining into a third string. This will be done in two steps. Firstly, we will construct
a state $\left|V\right>$ which satisfies
\begin{eqnarray}
\sum_{u=1}^3 p^I_{(u)}(\sigma)\left|V\right>&=&0\,,\qquad \sum_{u=1}^3 e(\alpha_u)x^I_{(u)}(\sigma)\left|V\right>=0\,,
\label{boscont}
\\
\sum_{u=1}^3 \lambda^A_{(u)}(\sigma)\left|V\right>&=&0\,,\qquad \sum_{u=1}^3 e(\alpha_u)\theta_{(u)A}(\sigma)\left|V\right>=0\,,\\
\sum_{u=1}^3 {\tilde\lambda}^A_{(u)}(\sigma)\left|V\right>&=&0\,,\qquad \sum_{u=1}^3 e(\alpha_u){\tilde\theta}_{(u)A}(\sigma)\left|V\right>=0\,.
\end{eqnarray}
These equations implement the functional delta function constraints~(\ref{deltfn}) at the 
Fock space level. The coordinates of the three strings shown in Figure~\ref{fig1} are parameterized by
\begin{eqnarray}
\s_1 & =&\s \qquad\quad\qquad 0\le\s\le\pi\a_1\,, \nonumber \\
\s_2 & =&\s-\pi\a_1 \quad\pi\a_1\le\s\le\pi(\a_1+\a_2)\,, \\
\s_3 & =&-\s \qquad\quad\qquad 0\le\s\le \pi(\a_1+\a_2)\,,
\end{eqnarray}
and $\a_1+\a_2+\a_3=0$ with $\a_3<0$. Further 
\begin{equation}
x^I_u(\sigma)=x^I_u(\sigma_u)\Theta_u\,,
\end{equation}
with similar expressions for the other fields, where
\begin{equation}
\Theta_1=\left\{\begin{array}{lcr}1 & \,\,\,& \sigma <\pi\alpha_1 \\ 0 & \,\,\,& \sigma>\pi\alpha_1\,,\end{array}\right.
\qquad
\Theta_2=\left\{\begin{array}{lcr}0 &\,\,\, & \sigma <\pi\alpha_1 \\ 1 &\,\,\, & \sigma>\pi\alpha_1\,,\end{array}\right.
\qquad\Theta_3=\Theta_1+\Theta_2=1\,.
\end{equation}
The above constraints can then be expanded in $\sigma$ Fourier modes to give the restrictions on $\left|V\right>$ in terms of
creation and annihilation operators\footnote{The
state $\left|V\right>$ can also be found by performing a functional integral over the functional delta function.} (see for example equation~(\ref{fermcont}) below).

Secondly, we will require that the interaction vertex together with the dynamical supercharges satify the supersymmetry algebra at ${\mc O}(g_{\text{s}})$. The full cubic vertex will be given by an exponential of creation
operators in the Fock spaces of the three strings, which enforces the kinematical constraints and a prefactor, polynomial in creation operators, which implements the dynamical constraints.
We conclude this section by re-interpreting the operator expressions for
the cubic interaction vertex and dynamical supercharges in a functional language. The reader
should be aware that using functional expressions directly in the plane-wave background is much more
suble than in flat spacetime. For example equation~(\ref{funnycom}) is no longer valid when the anti-commutator
does not act directly on $\left|V\right>$; in flat spacetime on the other hand there is no need
to place the commutator next to $\left|V\right>$ (see equation~(4.29b) of~\cite{gssft}). This
type of behaviour was already encountered in the closed string cubic vertex in the plane-wave 
background~\cite{ps}.

\subsection{The kinematical constraints}\label{sec31}

The bosonic continuity conditions, and mode expansions are very similar to the closed string ones. In particular 
the Neumann directions are like the positive modes of the closed
string and the Dirichlet directions like the negative modes. The bosonic part of the vertex is then
\begin{equation}
\left|V_{\mbox{bos}}\right>=\exp\left\{\frac{1}{2}\sum_{u,v=1}^3\left(\sum_{m,n}
a^{I\dagger}_{m(u)}{\bar N}^{uv}_{mn}a^{I\dagger}_{n(v)}
\right)\right\}\left|0\right>_{123}\,.
\end{equation}
where $|0\ra_{123}\equiv |0\ra_1\otimes|0\ra_2\otimes|0\ra_3$ with $|0\ra_u$ defined in 
equation~(\ref{notvacuum}) below. $N^{uv}_{mn}$ is the usual plane-wave
background Neumann matrix, whose explicit form (up to exponential corrections) has recently been 
found~\cite{hssv}. For $I$ Dirichlet the indices $m$ and $n$ run over the negative integers,
while for $I$ Neumann they run over the non-negative integers. It is easy to repeat the analysis 
of~\cite{ap} in order to confirm that $\left|V_{\mbox{bos}}\right>$ satisfies the continuity conditions
enforced by the functional delta function.

\noindent The continuity conditions for fermions can be recombined as
\begin{equation}\label{fermcont}
\sum_{u=1}^3 \lambda^A_{\pm(u)}(\sigma_u)\left|V\right>=0\,,\qquad 
\sum_{u=1}^3e(\alpha_u)\theta_{\pm(u)A}(\sigma_u)\left|V\right>=0\,,
\end{equation}
where $\lambda_{\pm(u)}$, $\theta_{\pm(u)}$ are defined in equations~(\ref{normlambda}) 
and~(\ref{normtheta}).
The functional delta functions can then be written as an infinite product of delta functions of the $\sin$ and $\cos$ modes. One may then perform the Gaussian integration to obtain the fermionic vertex. Instead we will write down the Fourier modes of the above constraints and solve for the vertex as an eigenstate of these equations. 
The continuity conditions~(\ref{fermcont}) reduce to 
\begin{equation}\label{fermcomp}
\sum_{u,n}X^{(u)}_{mn}\lambda_{n(u)}^A\left|V_{\mbox{ferm}}\right>=0\,,\qquad
\sum_{u,n}\alpha_uX^{(u)}_{mn}\theta_{n(u)A}\left|V_{\mbox{ferm}}\right>=0\,.
\end{equation}
The Fourier transform matrices $X^{(u)}_{mn}$ were defined in~\cite{ap}; we note here that they are block diagonal with the negative modes forming one block and the non-negative modes the other. Re-writing these in terms of
the creation and annihilation operators we get
\begin{equation}\label{modecond}
\begin{split}
\sum_{u=1}^3\sum_{n=1}^\infty\frac{1}{\alpha_u}(A^{(u)}CC^{-1/2}_{(u)})_{mn}(P_{n(u)C}{}^AR_{n(u)}^C
-P_{n(u)C}^{-1}{}^AR_{-n(u)}^C)\left|V_{\mbox{ferm}}\right>&=0\\
\left[\sum_{u=1}^3\sum_{n=1}^\infty(A^{(u)}C^{-1/2}_{(u)})_{mn}(P^{-1}_{n(u)C}{}^AR_{n(u)}^C
+P_{n(u)C}{}^AR_{-n(u)}^C)+B_m\Lambda^A\right]\left|V_{\mbox{ferm}}\right>&=0\\
\sum_{u=1}^3\sum_{n=1}^\infty(A^{(u)}CC^{-1/2}_{(u)})_{mn}(P^{-1}_{n(u)A}{}^CR_{n(u)C}
-P_{n(u)A}{}^CR_{-n(u)C})\left|V_{\mbox{ferm}}\right>&=0\\
\left[\sum_{u=1}^3\sum_{n=1}^\infty\alpha_u(A^{(u)}C^{-1/2}_{(u)})_{mn}(P_{n(u)A}{}^CR_{n(u)C}
+P^{-1}_{n(u)A}{}^CR_{-n(u)C})-\alpha B_m\Theta_A\right]\left|V_{\mbox{ferm}}\right>&=0\,.
\end{split}
\end{equation}
where 
\begin{equation}
\alpha\equiv\alpha_1\alpha_2\alpha_3\,,
\end{equation}
and the matrices $A^{(u)}$, $C$, $C_{(u)}$ and vector $B$ are the same as in~\cite{ps} and
\begin{equation}
\alpha_3\Theta_A\equiv\theta_{0(1)A}-\theta_{0(2)A}\,,\qquad
\Lambda^A=\alpha_1\lambda_{0(2)}^A-\alpha_2\lambda_{0(1)}^A\,.
\end{equation}
In appendix~\ref{appa} we show that these are satisfied by
\begin{equation}\label{fermexp}
\left|V_{\mbox{ferm}}\right>=\exp\left\{\sum_{u,v=1}^3\sum_{m,n=1}^\infty
R_{-m(u)}^AQ^{uv}_{mn}{}_A{}^BR_{-n(v)B}-\sum_{u=1}^3\sum_{m=1}^\infty R^A_{-m(u)}Q^u_m{}_A{}^B\Theta_B
\right\}\left|V^0_{\mbox{ferm}}\right>\,,
\end{equation}
with
\begin{equation}
\left|V^0_{\mbox{ferm}}\right>=\prod_{A=1}^4\left[\sum_{u=1}^3\alpha_u\theta_{0(u)A}\right]
\left|0\right>_{123}\,,
\end{equation}
enforcing the zero-mode constraints. Here
\begin{equation}\label{theqs}
\begin{split}
Q^u_{nA}{}^B&=\frac{\alpha/\alpha_u}{1-4\mu\alpha K}(1-2\mu\alpha K(1+i\Omega\Pi))_A{}^C
(C_{(u)}^{1/2}P_{(u)C}{}^BC^{1/2}{\bar N}^u)_n\,,\\
Q^{uv}_{mnA}{}^B&=\frac{\alpha_v}{\alpha_u}
(P_{(u)A}^{-1}{}^CU_{(u)}C^{1/2}{\bar N}^{uv}C^{-1/2}U_{(v)}P_{(v)C}^{-1}{}^B)_{mn}\,.
\end{split}
\end{equation}
The state $\left|0\right>_{123}\equiv |0\ra_1\otimes|0\ra_2\otimes|0\ra_3$ on which the vertex is built is defined as
\begin{eqnarray}
a^I_{n(u)}|0\ra_u=0\,,\qquad R_{m(u)}^A|0\ra_u&=&0\,,\qquad R_{m(u)A}|0\ra_u=0\,,\quad m > 0\nonumber\\
\lambda_{0(u)}^A|0\ra_u&=&0\,,\label{notvacuum}
\end{eqnarray}
with $n$ negative (non-negative) for $I$ Dirichlet (Neumann).

\noindent In summary, the part of the cubic contribution to the dynamical generators satisfying the
kinematic constraints is
\begin{equation}
|V\ra\equiv \left|V_{\mbox{ferm}}\right>\left|V_{\mbox{bos}}\right>\d\left(\sum_{r=1}^3\a_r\right)\,.
\end{equation}

\subsection{The dynamical constraints}\label{sec32}

In interacting lightcone string field theory $H$, $q^-_A$ and $q^{-A}$ receive $g_s$ corrections, while
operators such as $J$, $P$ and $q^+$ do not. As a result, the superalgebra relation~(\ref{sualg}) at ${\mc O}(g_s)$ implies
\begin{equation}
\sum_{u=1}^3q^-_{A(u)}\left|Q^{-B}\right>+
\sum_{u=1}^3q^{-B}_{(u)}\left|Q^-_A\right>=2\delta_A^B\left|H\right>\,.
\end{equation}
This relation places tight constraints on the form of the dynamical supercharges and cubic interaction vertex. These dynamical constraints are solved by 
introducing a prefactor~\cite{gs,gsb}, polynomial in creation operators, in front of $|V\ra$. The constituents of
this polynomial prefactor have to commute with the dynamical constraints
\begin{eqnarray}\label{bospre}
\bigl[\,\sum_{u=1}^3p_u(\s_u),{\mc P}\bigr]&=&0
=\bigl[\,\sum_{u=1}^3e(\a_u)x_u(\s_u),\mc{P}\bigr]\,,\\
\label{ferpre}
\bigl\{\,\sum_{u=1}^3\l_u(\s_u),{\mc P}\bigr\}&=&0=
\bigl\{\,\sum_{u=1}^3e(\a_u)\vt_u(\s_u),\mc{P}\bigr\}\,.
\end{eqnarray}
The bosonic constituents are again similar to the closed string case, and for the Neumann directions they give
\begin{equation}\label{k+mode}
{\mc K}^r\equiv{\mc K}^r_0+{\mc K}^r_+\equiv
\mathbb{P}^r-i\m\frac{\a}{\a'}\mathbb{R}^r
+\sum_{u=1}^3\sum_{n=1}^{\infty}F_{n(u)}a^{\dg r}_{n(u)}\,,
\end{equation}
where
\begin{equation}
\mathbb{P}^r\equiv\a_1p^r_{0(2)}-\a_2p^r_{0(1)}\,,\qquad
\a_3\mathbb{R}^r\equiv x^r_{0(1)}-x^r_{0(2)}\,,\qquad [\mathbb{R}^r,\mathbb{P}^s]=i\delta^{rs}
\end{equation}
and in terms of the zero-mode creation oscillators
\begin{equation}
\mathbb{P}^r-i\m\frac{\a}{\a'}\mathbb{R}^r=
\sqrt{\frac{2}{\a'}}\sqrt{\m\a_1\a_2}
\bigl(\sqrt{\a_1}a_{0(2)}^{\dg r}-\sqrt{\a_2}a_{0(1)}^{\dg r}\bigr)\,.
\end{equation}
In the Dirichlet directions the bosonic prefactor will be built out of
\begin{equation}\label{k-mode}
{\mc K}^{r^\prime,\,L,\,R}\equiv\sum_{u=1}^3\sum_{n=1}^{\infty}F_{-n(u)}a^{\dg r^\prime,\,L,\,R}_{-n(u)}\,,
\end{equation}
where $F_{(u)}$ is the same as in the closed string.

\noindent The solution involving subscripted fermionic creation operators  is
\begin{equation}\label{ymode}
{\mc Y}_A\equiv-\frac{\alpha}{\sqrt{\a^\prime}}\Theta_A
+\sum_{u=1}^3\sum_{n=1}^{\infty}G_{n(u)A}{}^BR_{-n(u)B}\,,
\end{equation}
where 
\begin{equation}
G_{n(u)A}{}^B=-\frac{\alpha/\sqrt{\alpha^\prime}}{1-4\mu\alpha K}(1-2\mu\alpha K(1-i\Omega\Pi))_A{}^C
(P^{-1}_{(u)nC}{}^BC^{1/2}_{(u)}C^{1/2}{\bar N}^u)_n\,.
\end{equation}
As in flat space~\cite{gs,gsb}, it turns out that the prefactors
do not involve superscripted fermionic creation oscillators.

\noindent Given the above, the simplest ansatz for the dynamical supercharges is
\begin{eqnarray}\label{qbot}
\left|Q^-{}_A\right>&=&Y_A\left|V\right>\,,\\\label{qtop}
\left|Q^{-A}\right>&=&k\epsilon^{ABCD}Y_{B}Y_{C}Y_{D}\left|V\right>\,,
\end{eqnarray}
with $k$ a constant to be determined and
\begin{equation} 
Y_A\equiv(1-4\m\a K)^{-1/2}(1-2\m\a K(1+i\Omega\Pi))_A{}^B{\mc Y}_B\,.
\end{equation}
One can show that 
\begin{eqnarray}
\sum_{u=1}^3q_{(u)}^{-A}\left|V\right>
&=&-\sqrt{2\alpha^\prime}\frac{\alpha^\prime}{\alpha}
({\mc K}^r\rho^r+{\mc K}^{r^\prime}\rho^{r^\prime})^{AB}(1-2\mu\alpha K(1+i\Omega\Pi))_B{}^C{\mc Y}_C
\left|V\right>\,,\\
\sum_{u=1}^3q^-_{(u)A}\left|V\right>
&=&-2\sqrt{\alpha^\prime}\frac{\alpha^\prime}{\alpha}
{\mc K}^R(1-2\mu\alpha K(1+i\Omega\Pi))_A{}^B{\mc Y}_B
\left|V\right>\,,
\end{eqnarray}
and
\begin{eqnarray}
\left\{\sum_{u=1}^3q_{(u)}^{-A},{\mc Y}_B\right\}\left|V\right>&=&2\sqrt{\alpha^\prime}{\mc K}^L
(1-2\mu\alpha K(1-i\Omega\Pi))_B{}^A\left|V\right>\,,\\ \label{funnycom}
\left\{\sum_{u=1}^3q_{(u)A}^-,{\mc Y}_B\right\}\left|V\right>&=&-\sqrt{2\alpha^\prime}
({\mc K}^r\rho^r+{\mc K}^{r^\prime}\rho^{r^\prime})_{AC}(1-2\mu\alpha K(1-i\Omega\Pi))_B{}^C
\left|V\right>\,.
\end{eqnarray}
Using these one is led to the solution
\begin{equation}\label{hint}
\left|H\right>=(1-4\mu\alpha K)^{1/2}\Bigl[\sqrt{\alpha^\prime}{\mc K}^L+
\sqrt{\frac{\alpha^\prime}{2}}\frac{\alpha^\prime}{2\alpha}{\mc K}^i\rho^{iCD}Y_CY_D
+\frac{(\alpha^\prime)^{5/2}}{24\alpha^2}{\mc K}^R\epsilon^{CDEF}Y_CY_DY_EY_F\Bigr]\left|V\right>\,,
\end{equation}
and 
\begin{equation}
k=-\frac{\alpha^\prime}{6\alpha}\,.
\end{equation}
It is important to note that as in the case of the closed string in the plane-wave background, the overall normalisation of the dynamical supercharges, and hence also the cubic vertex cannot be fixed in this way.\footnote{The normalisation may well be a $\mu$ dependent quantity. Since we have constructed the vertex by enforcing continuity conditions, rather then by performing a functional integral, we have already dropped a normalisation coming from the mismatch between the fermionic and bosonic determinants.
The non-zero modes' determinants cancel between bosons and fermions. On the other hand the bosonic zero-modes contribute a factor $(4\mu\alpha_1\alpha_2/\alpha'\pi^3\alpha_3)^{(p-1)/4}$ for a D$p$-brane
while the fermionic zero modes do not contribute at all. It would be tempting then to speculate that the relative normalisations of $\left|H\right>$ between a D$p$- and a D$p^\prime$-brane would be
$(4\mu\alpha_1\alpha_2/\alpha'\pi^3\alpha_3)^{(p-p^\prime)/4}$.}

\subsection{Functional interpretation}\label{sec33}

The bosonic constituents of the prefactor can be obtained via the operators~\cite{gs,gsb}
\begin{equation}
\begin{split}
\p X^I(\s) & =
4\pi\frac{\sqrt{-\a}}{\a'}(\pi\a_1-\s)^{1/2}\bigl(\partial_\sigma x^I_1(\s)+\partial_\sigma x^I_1(-\s)\bigr)\,,\\
\label{p}
P^I(\s) &=-2\pi\sqrt{-\a}(\pi\a_1-\s)^{1/2}\bigl(p^I_1(\s)+p^I_1(-\s)\bigr)\,.
\end{split}
\end{equation}
Acting on the exponential part of the vertex these satisfy
\begin{equation}\label{pp}
\begin{split}
\lim_{\s\to\pi\a_1}K^r(\s)|V\ra & \equiv
\lim_{\s\to\pi\a_1}P^r(\s)=f(\m){\mc K}^r|V\ra\,,\\
\lim_{\s\to\pi\a_1}K^{r^\prime,L,R}(\s)|V\ra & \equiv
\lim_{\s\to\pi\a_1}\frac{1}{4\pi}\p X^{r^\prime,L,R}(\s)=f(\m){\mc K}^{r^\prime,L,R}|V\ra\,.
\end{split}
\end{equation}
Here we defined
\begin{equation}\label{fmu}
f(\m)\equiv-2\frac{\sqrt{-\a}}{\a_1}\lim_{e\to0}\e^{1/2}\sum_{n=1}^{\infty}(-1)^nn\cos(n\e/\a_1)\bar{N}^1_n=(1-4\mu\alpha K)^{1/2}
\end{equation}
where the last equality was conjectured in~\cite{ps} and proved in~\cite{hssv}. For the fermionic constituent of the prefactor one considers~\cite{gs,gsb}
\begin{equation}
Y(\s)=-2\pi\frac{\sqrt{-2\a}}{\sqrt{\a'}}(\pi\a_1-\s)^{1/2}\bigl(\theta_1(\s)+\theta_1(-\s)\bigr)
\end{equation}
which satisfies
\begin{equation}
\lim_{\s\to\pi\a_1}Y(\s)|V\ra=f(\m)(1-4\m\a K)^{-1}(1-2\m\a K(1+i\Omega\Pi)){\mc Y}|V\ra=Y|V\ra
\,.
\end{equation}
The oscilator expressions for the dynamical supercharges may then be replaced by functional integrals.
For example
\begin{equation}
Q^{-}_{3A}  = g_s\int d\m_3Y_A\tr(\Phi(1)\Phi(2)\Phi(3))\,,
\end{equation}
where the trace is over Chan-Patton factors. Here
\begin{eqnarray}
d\m_3&\equiv& \left(\prod_{u=1}^3d\a_uD^4\l_{+u}(\s)D^4\l_{-u}(\s)D^8p_u(\s)\right)\nonumber \\
&&\qquad\qquad\times\,\,\,
\d\bigl(\sum_v\a_v\bigr)\D^4\bigl[\sum_v\l_{+v}(\s)\bigr]\D^4\bigl[\sum_v\l_{-v}(\s)\bigr]
\D^8\bigl[\sum_vp_v(\s)\bigr]
\end{eqnarray}
is the functional expression leading to the exponential part of the vertex \cite{gs,gsb}. Similar expressions can be written down for $Q_3^A$ and $H_3$ with $Y_A$ replaced by corresponding expressions from equations~(\ref{qtop}) and~(\ref{hint}).

\section{Conclusion}\label{sec4}

We have constructed the cubic interaction vertex and dynamical supercharges for D$_-$-branes in the plane-wave background. The expressions were obtained in terms of oscilator expressions, allowing for immediate computation of three-point functions of open strings ending on such D$_-$-branes. They have also been reinterpreted in the functional language. In the above construction the $SU(4)$ formalism has been used. The $SU(4)$ language allows for a unified treatment of all D$_-$-branes even though they do not preserve an $SU(4)$ symmetry. The situation is analogous to the closed string vertex in the plane-wave background, where $SO(8)$ notation is used, with the theory preserving only an $SO(4)\times SO(4)$ subgroup. While the explicit expressions given here have been for D-branes placed at the origin in transverse space, as was discussed in the introduction, it is straightforward to generalise this to other D$_-$-branes. The oscilator formulas have been constructed on the state $\left|0\right>_{123}$ rather than on the open string vacuum $\left|v\right>_{123}$. This too is similar to the closed string vertex. It may be interesting to investigate if the open string cubic vertex could be also constructed on $\left|v\right>_{123}$, in the spirit of~\cite{ov}. Finally, we note that it is of course possible to add terms of the form $\Sigma_u q^-_{(u)}\left|V\right>$ to the dynamical supercharges and 
$\Sigma_u H_{2(u)}\left|V\right>$ to the cubic hamiltonian as has been recently suggested by~\cite{divecchia}.

The D$_-$-branes in the plane-wave background are expected to be dual to the BMN-like limit of certain defect theories. It would be interesting to compare the open-string three-point functions which can be obtained using the three-point vertex computed here with the corresponding gauge theory results. It would also be interesting to extend the string bit model~\cite{bits} to include open strings.

\vskip1cm
\section*{Acknowledgements}
I would like to thank Ari Pankiewicz for discussions and comments on the manuscript.
I am greatful to Luis Iba\~nez, Angel Uranga and the Physics group at Universidad Autónoma de Madrid 
for their hospitality during the final stages of this project.
This work was supported by FOM, the Dutch Foundation for Fundamental Research on Matter.

\appendix
\section{Dynamical fermionic constraints}\label{appa}

In this appendix we show that the continuity conditions~(\ref{fermcomp}) are solved by~(\ref{fermexp}). Note first that given the ansatz~(\ref{fermexp}) equations~(\ref{modecond}) become
\begin{eqnarray}
\sum_{u,n}A^{(u)}_{mn}C_nC^{-1/2}_{n(u)}
P^{-1}_{n(u)A}{}^CQ^u_{nC}{}^B&=&0\,,\\
-\alpha B_m\delta_A^B+\sum_{u,n}\alpha_uA^{(u)}_{mn}C^{-1/2}_{(u)n}
P_{(u)nA}{}^CQ^u_n{}_C{}^B&=&0\,,\\
(A^{(v)}CC^{-1/2}_{(v)}P_{(v)}{}_A{}^B)_{mn}-\sum_{u=1}^3
(A^{(u)}CC^{-1/2}_{(u)}P^{-1}_{(u)A}{}^CQ^{uv}{}_C{}^B)_{mn}&=&0\,,\\
\alpha_v(A^{(v)}C_{(v)}^{-1/2}P_{(v)}^{-1}{}_A{}^B)_{mn}+
\sum_{u}\alpha_u(A^{(u)}C^{-1/2}_{(u)}P_{(u)}{}_A{}^CQ^{uv}{}_C{}^B)_{mn}&=&0\,,\\
\frac{1}{\alpha_v}(P^{-1}_{(v)}{}_B{}^ACC^{-1/2}_{(v)}A^{(v)})_{mn}+
\sum_{u}\frac{1}{\alpha_u}(Q^{vu}{}_B{}^CP_{(u)C}{}^AC^{-1/2}_{(u)}CA^{(u)})_{mn}&=&0\,,\\
2Q^v_m{}_B{}^AB_n
+(P_{(v)}{}_B{}^AC^{-1/2}_{(v)}A^{(v)})_{mn}
-\sum_u(Q^{vu}_B{}^CP_{(u)}^{-1}{}_C{}^AC^{-1/2}_{(u)}A^{(u)})_{mn}
&=&0\,.
\end{eqnarray}
In order to verify that the solutions are given in equations~(\ref{theqs}) one needs to generalise some of the fermionic identities in~\cite{ap}. For example
\begin{equation}
P_{(u)A}^{-2}{}^CU_{(u)}{\bar N}^{uv}U_{(v)}P_{(u)C}^{-2}{}^B=\delta_A^B{\bar N}^{uv}
+\mu\alpha(1-i\Omega\Pi)_A{}^BC^{1/2}_{(u)}{\bar N}^u(C^{1/2}_{(v)}{\bar N}^v)^T
\end{equation}
The above constraints are then shown to be satisfied identically using the bosonic identities in~\cite{sch,ap}.


\end{document}